\documentclass[12pt]{article}
 \usepackage{graphicx,amsfonts,amsmath,amssymb,latexsym}
 \renewcommand{\title}[1] {%
 \begingroup\begin{center}\vspace{0.0cm}\bf\Large
 \addtolength{\baselineskip}{1mm} #1 \end{center}\endgroup}

 \renewcommand{\author}[1] {%
 \begingroup\begin{center}\vspace{0.2cm}\bf #1 \vspace{0.2cm}
 \end{center}\endgroup}

 \newcommand{\address}[1] {%
 \begingroup\begin{center} #1 \end{center}\endgroup}

 \newcommand\bz{\bar{z}}
 
 \newcommand\Zb{\mathbb{Z}}
 \newcommand\Rb{\mathbb{R}}
 \newcommand\Cb{\mathbb{C}}

 \newcommand\ben{\begin{equation*}}
 \newcommand\ebn{\end{equation*}}
 \newcommand\be{\begin{equation}}
 \newcommand\eb{\end{equation}}

 \numberwithin{equation}{section}

\begin{document}
 \title{Aharonov-Bohm effect on the Poincar\'e disk}
 \author{Oleg Lisovyy}
 \address{
 Laboratoire de Math\'ematiques et Physique Th\'eorique CNRS/UMR 6083,\\
 Universit\'e de Tours,
   Parc de Grandmont, 37200 Tours, France}
 \begin{abstract}
 We consider formal quantum Hamiltonian of a charged particle on
 the Poincar\'e disk in the presence of an Aharonov-Bohm magnetic vortex
 and a uniform magnetic field. It is shown that this Hamiltonian
 admits a four-parameter family of self-adjoint extensions. Its resolvent
 and the density of states are calculated for natural values of the extension
 parameters.
 \end{abstract}

 \section{Introduction}
 Quantum dynamics on the Poincar\'e disk has long been a subject
 of theoretical interest, mainly because of the insights its study
 provides into the theory of quantum chaos. Analyzed examples
 include, for instance, the free motion under the action of
 constant magnetic fields \cite{comtet_houston, comtet,
 grosche_h}, the Kepler problem \cite{grosche_c},
 the scattering by the Aharonov-Bohm (AB) \cite{grosche,kuperin} and
  Aharonov-Bohm-Coulomb \cite{nouicer} potentials,
  the study of point
 interactions \cite{aeg, bruning} and quantum Hall effect~\cite{bgm}.

 In the present paper, we consider the Hamiltonian of a charged
 spinless particle moving on the hyperbolic disk, pierced by an AB
 flux, in the presence of a uniform magnetic field. First part of
 this work is rather standard: we determine the admissible
 boundary conditions on the wave functions, using Krein's theory of
 self-adjoint extensions (SAEs) \cite{albeverio}. It turns out that in the most
 general case the formal Hamiltonian has deficiency indices $(2,2)$ and
 thus admits a four-parameter
 family of SAEs. Let us remark that similar results on the
 plane
 have been found in \cite{adami, stovicek1} in the case of
 zero magnetic field,  and in \cite{stovicek2} for non-zero
 fields;   SAEs of the Dirac Hamiltonian on the plane
 have been studied in \cite{desousa,falomir}.

  The rest of this paper is
 devoted to the study of a particular extension, corresponding to
 the choice of regular boundary conditions at the  position of the AB flux.
 We start by constructing certain integral representations for common
 eigenstates of this Hamiltonian and the angular momentum
 operator. These representations then allow to sum up the contributions coming from
 different angular momenta  to the resolvent kernel, and to evaluate this kernel and
 the density of states in
 a closed form.

 The material is organized as follows. In Section~2, we introduce
 basic notations and study elementary solutions of the radial
 Schr\"odinger equation on the Poincar\'e disk. Self-adjointness
 of the full AB Hamiltonian is discussed in Section~3.
 In  Section~4 we find a compact expression for the resolvent of
 the regular extension (formulas \ref{ans}),
 (\ref{g0ans})--(\ref{vrr})). These relations represent
 the main result of the present work. The density of states, induced by the AB flux
 in the whole hyperbolic space (see (\ref{finresd})--(\ref{rhocont})), is obtained in Section~5.
 Some technical results are
 relegated to the appendices.

 \section{Free Hamiltonian on the Poincar\'e disk}
 \subsection{Basic formulas}
 Let us identify the Poincar\'e disk $D=SU(1,1)/SO(2)$ with the interior
 of the unit circle $|z|^2<1$ in the complex plane, equipped with the metric
 \be\label{pmetric}
 ds^2=g_{z\bz}\,dz\, d\bz=R^2\frac{dz\,d\bz}{\left(1-|z|^2\right)^2}
 \eb
 of constant Gaussian curvature $-4/R^{2}$.
 We consider a spinless particle moving on the disk and interacting with
 a magnetic field. The latter can be introduced as a connection
 1-form
 \ben
 \mathcal{A}=A_z\,dz+A_{\bz}\,d\bz
 \ebn
 on the trivial $U(1)$-bundle over $D$. Quantum dynamics of a
 particle of unit charge is described by the Hamiltonian
 \be\label{hamgen}
 \hat{H}=-\frac{2}{g_{z\bz}}\,\left\{D_z,D_{\bz}\right\},
 \eb
 where $D_z=\partial_z+iA_z$ and $D_{\bz}=\partial_{\bz}+iA_{\bz}$
 are the usual covariant derivatives. To unburden formulas, we put
 the particle mass equal to $1/2$ and $\hbar =c=1$ throughout the paper.

 In the remainder of the present
 section, the following vector potential is considered:
 \be\label{umf}
 \mathcal{A}^{(B)}=-\frac{i{BR}^{\,2}}{4}\,\frac{\bz\, dz -
 z\,d\bz}{1-|z|^2}\,.
 \eb
 It generates a curvature 2-form $\mathcal{F}^{(B)}$, proportional to
 the invariant volume measure $\displaystyle d\mu=\frac{i}{2}\,g_{z\bz}\,dz\wedge
 d\bz$. Indeed, we have
 \ben
 \mathcal{F}^{(B)}=d\mathcal{A}^{(B)}=Bd\mu\,.
 \ebn
 Therefore, the potential (\ref{umf}) describes a uniform
 magnetic field of intensity $B$. Introducing polar coordinates
 $z=re^{i\varphi}$, $\bz=re^{-i\varphi}$, one can write the
 corresponding Hamiltonian as
 \be\label{hamumf}
 \hat{H}^{(B)}=-\frac{\left(1-r^2\right)^2}{R^2}\left\{\partial_{rr}+\frac{1}{r}\,\partial_r
 +\frac{1}{r^2}\,\partial_{\varphi\varphi}+\frac{iBR^2}{1-r^2}\,\partial_{\varphi}-
 \frac{B^2R^4}{4\left(1-r^2\right)^2}\,r^2\right\}.
 \eb
 Note that the domain of $\hat{H}^{(B)}$ is not yet specified. It
 will be fixed in the next section by the requirement for the
 Hamiltonian to be a self-adjoint operator. According to Stone's
 theorem, this condition ensures the existence of consistent
 dynamics.

 \subsection{Radial Hamiltonians}
 Formal Hamiltonian $\hat{H}^{(B)}$ commutes with the angular
 momentum operator $\hat{L}=-i\partial_{\varphi}$. Therefore, it
 leaves invariant the eigenspaces of $\hat{L}$, spanned by the
 functions $w_l(r)e^{il\varphi}$ ($l\in\Zb$). Being restricted to
 the eigenspace of $\hat{L}$, characterized by the angular
 momentum $l$, the Hamiltonian acts as follows:
 \ben
 w_l(r)\mapsto \hat{H}_l\, w_l(r),
 \ebn
 \be\label{hamufml}
 \hat{H}_l=-\frac{\left(1-r^2\right)^2}{R^2}\left\{\partial_{rr}+\frac{1}{r}\,\partial_r
 -\frac{l^2}{r^2}-\frac{4b\,l}{1-r^2}-
 \frac{4b^2r^2}{\left(1-r^2\right)^2}\right\}.
 \eb
 Here we have introduced instead of $B$ a dimensionless parameter
 $b=BR^2/4$.

 It will be useful for us to let the parameter $l$ to take on not
 only integer, but also arbitrary real values, and to study in
 some detail the properties of solutions of the radial
 Schr\"odinger equation
 \be\label{sche}
 \left( \hat{H}_l-k^2\right)w_l=0.
 \eb
 In what follows it will be always assumed that $k^2\in \Cb\backslash\Rb^+\cup\{0\}$.
 It is also convenient to introduce instead of $r$ a new
 variable $t=r^2$.

 We are interested in the solutions of
 (\ref{sche}) leading to square integrable (with the measure
 $d\mu$) functions on $D$. These solutions should be then square
 integrable on the open interval $I=(0,1)$ with the measure
 $\displaystyle d\mu_t=\frac{R^2dt}{2(1-t)^2}$. For each $l\in\Rb$
 there exists only one solution of (\ref{sche}), which is square
 integrable in the neighbourhood of the point $t=1$. Its explicit
 form is
 \begin{eqnarray}\label{sqis}
  w_l^{(I)}(t)&=& t^{-l/2}\left(1-t\right)^{\chi}
  {}_2F_1\left(
  \chi-b,\chi+b-l,2\chi,1-t\right)=\\
  \nonumber &=& t^{\,l/2}\;\;\left(1-t\right)^{\chi}
  {}_2F_1\left(
  \chi+b,\chi-b+l,2\chi,1-t\right),
 \end{eqnarray}
 where
 \ben
 \chi=\frac{1+\sqrt{1+4 b^2-k^2R^2}}{2}
 \ebn
 and ${}_2F_1(\alpha,\beta,\gamma,z)$ denotes Gauss hypergeometric
 function. The branches of square roots are defined so that
 they take on real positive values for purely imaginary $k$.

  Similarly, for each $l\in(-\infty,-1]\cup[1,\infty)$
 there is only one solution of (\ref{sche}),
 which is square integrable  with respect to $d\mu_t$
 near the point $t=0$. The  form of this solution depends on whether
 $l\geq1$ or $l\leq-1$. In the first case, i.~e. for $l\geq 1$, it is given by
 \be\label{ausol1}
  w_l^{(II,+)}(t)= t^{l/2}\left(1-t\right)^{\chi}
  {}_2F_1\left(
  \chi+b,\chi-b+l,1+l,t\right)
 \eb
 while for $l\leq-1$ this solution is written as follows:
 \be\label{ausol2}
   w_l^{(II,-)}(t)= t^{-l/2}\left(1-t\right)^{\chi}
  {}_2F_1\left(
  \chi-b,\chi+b-l,1-l,t\right).
 \eb
 Note  that for $|l|<1$ both functions $w_{l}^{(II,\pm)}(t)$ are
 square integrable in the vicinity of the point $t=0$ and solve the radial Schr\"odinger
 equation (\ref{sche}). These solutions are linearly independent except for
 $l=0$. However, in the latter case the equation (\ref{sche})  still admits
 two distinct
 solutions that are square integrable as
 $t\rightarrow0$:
 \begin{eqnarray*}
 w_0^{(II)}(t)=\left(1-t\right)^{\chi}u(t),\qquad
 \tilde{w}_0^{(II)}(t)=\left(1-t\right)^{\chi}v(t),
 \end{eqnarray*}
 where $u$ and $v$ are any two linearly independent solutions of
 the hypergeometric equation with parameters $\alpha=\chi+b$,
 $\beta=\chi-b$, $\gamma=1$ (one can choose them, for instance,
 according to the formulas 15.5.16 and 15.5.17 of \cite{abst}).

  Let us now show that the solutions $w_{l}^{(I)}(t)$ and  $w_{l}^{(II,+)}(t)$
 are linearly independent for $l>-1$, and the solutions $w_{l}^{(I)}(t)$ and  $w_{l}^{(II,-)}(t)$
 are linearly independent for $l<1$. This can be done by an explicit
 computation of their Wronskian
 \ben
 W(f_1,f_2)=f_1\cdot\partial_tf_2-\partial_tf_1\cdot f_2\,.
 \ebn
 Namely, using
 the connection and analytic continuation formulas for hypergeometric functions
 \cite{abst}, one obtains
 \be
 \label{wronsk120}
 W\left(w_{l}^{(I)}(t),w_{l}^{(II,\pm)}(t)\right)=
 \left(t\,C^{\pm}_{k,l}\right)^{-1},
 \eb
 with
 \be \label{wronsk121}
 C^{\pm}_{k,l}=
 \frac{\Gamma(\chi\pm b)\Gamma(\chi\mp b\pm l)}{\Gamma(2\chi)\Gamma(1\pm
 l)}\,.
 \eb
 Therefore, for $k^2\in\Cb\backslash\Rb^+\cup\{0\}$ and $|l|\geq1$
 the equation (\ref{sche}) has
 no square integrable solutions (with the measure $d\mu_t$)
 on the whole interval $I$. This is true, in particular, for all
 radial Hamiltonians $\hat{H}_{l\in\Zb}$ of the free particle in a uniform
 magnetic field, except for the $s$-wave Hamiltonian $\hat{H}_0$.
 In the case $|l|<1$ the equation
 (\ref{sche}) has exactly one square integrable solution,
 given by the formula (\ref{sqis}).

  Let us now restrict the domain of  $\hat{H}_l$ to
 $\mathcal{D}(\hat{H}_l)={C}_0^{\infty}(I)$, i.~e. to smooth compactly supported functions.
 Then the above remarks imply that
 \begin{itemize}
 \item $\hat{H}_l$ is essentially self-adjoint for $|l|\geq1$,
 \item for $|l|<1$ the operator $\hat{H}_l$ has
 deficiency indices $(1,1)$ and thus admits a one-parameter family of
 self-adjoint extensions (SAEs).
 \end{itemize}
  Different extensions $\hat{H}_l^{(\gamma)}$
 ($|l|<1$) are in one-to-one correspondence with the isometries
 between the deficiency subspaces $\mathcal{K}^{\pm}_l=\mathrm{ker}
 \left(\hat{H}_l\mp i\varepsilon\right)$, where $\varepsilon\in\Rb^+$ may be chosen
 arbitrarily.
 They can be labeled by a real parameter $\gamma\in[0,2\pi)$ and
 characterized by the domains
 \ben
 \mathcal{D}(\hat{H}_l^{(\gamma)})=\left\{f+c\left(w_l^++e^{i\gamma }w_l^-\right)\,|\,
 f\in {C}_0^{\infty}(I),c\in\Cb\right\},
 \ebn
 where the functions $w^{\pm}_l(t)$ may be chosen as follows:
 \be\label{wpm}
 w^{\pm}_l(t)=w^{(I)}_{l}(t)\Bigl|_{k^2=\pm i\varepsilon}\Bigr. .
 \eb
  \textbf{Remark}. For a particular value of
 $\gamma$ the domain $\mathcal{D}(\hat{H}_l^{(\gamma)})$ is composed of
 functions, regular at $t=0$. The corresponding SAE of $\hat{H}_l$
 will be denoted by $\hat{H}_l^{\text{reg}}$.

  \subsection{Resolvent}
 The kernel $G_{k,l}(t,t')$ of the resolvent of the radial Hamiltonian
 $\hat{H}_l$ satisfies the equation
 \be\label{gfeq}
 \Bigl(\hat{H}_l(t)-k^2\Bigr)G_{k,l}(t,t')=
 \frac{2(1-t)^2}{R^2}\,\delta(t-t')\,.
 \eb
 It basically means that if
 $ \left(\hat{H}_l-k^2\right)u=v$ for some $u\in\mathcal{D}(\hat{H}_l) $, then
 \ben
 u(t)=\int\nolimits_I G_{k,l}(t,t')v(t')\,d\mu_{t'}.
 \ebn
 In order to find the solution of the equation (\ref{gfeq}), consider the
 following ansatz:
 \be\label{gfans}
 G_{k,l}(t,t')=
  \begin{cases}
    \tilde{C}^{\,\pm}_{k,l}\;w^{(II,\pm)}_l(t)\; w^{(I)}_l(t') & \text{for}\;\; 0<t<t'<1, \\
    \tilde{C}^{\,\pm}_{k,l}\;w^{(I)}_l(t)\; w^{(II,\pm)}_l(t') & \text{for}\;\;
    0<t'<t<1,
  \end{cases}
 \eb
 where the signs ``$+$'' and ``$-$'' should be chosen for $l\geq0$ and  $l<0$,
 correspondingly.
 It is clear that the function, defined by (\ref{gfans}),
 solves the equation (\ref{gfeq}) for $t\neq t'$ and satisfies the
 boundary conditions of square integrability at the points $t=0$ and
 $t=1$. (In the case $|l|<1$ the requirement of square integrability at the
 boundary points is not sufficient to make the operator $\hat{H}_l$
 self-adjoint; however, for such $l$, the ansatz (\ref{gfans}) also satisfies the
 regularity condition at $t=0$ and thus corresponds to the resolvent
 of the extension $\hat{H}_l^{\text{reg}}$).

 Taking into account the explicit form of the operator $\hat{H}_l$,
 one may show that the required singular behaviour of the Green function at
 the point $t=t'$ is guaranteed provided the condition
 \ben
 \partial_{t'}G_{k,l}(t',t)\Bigl|_{t-0}^{t+0}\Bigr.=-\frac{1}{2t}
 \ebn
 holds. Using (\ref{gfans}), one can rewrite this condition as
 \ben
 2t\, \tilde{C}^{\,\pm}_{k,l}\cdot
 W\left(w_{l}^{(I)}(t),w_{l}^{(II,\pm)}(t)\right)=1\,.
 \ebn
 It follows from  (\ref{wronsk120}) that
 the last relation is satisfied if we choose
 $\tilde{C}_{k,l}^{\,\pm}=C_{k,l}^{\,\pm}/2$.
 Substituting this expression into (\ref{gfans}), one finds a
 representation for the radial Green functions $G_{k,l}(t,t')$.

  \section{Hamiltonian in the presence of a magnetic vortex}
 \subsection{Radial Hamiltonians}
 Let us now add to the Hamiltonian the field of an Aharonov-Bohm magnetic
 flux $\Phi=2\pi\nu$, centered at $z=0$:
 \be\label{abfield}
 \mathcal{A}^{(v)}=-\frac{i\nu}{2}\left(\frac{dz}{z}-\frac{d\bz}{\bz}\right).
 \eb
 This choice of the flux position involves no loss of generality, since
 we have a well-known transitive $SU(1,1)$-action on $D$, which preserves the
 metric (\ref{pmetric}):
 \be\label{su11}
 z\mapsto z_g=\frac{\alpha z+\beta}{\bar{\beta}z+\bar{\alpha}}\,,\qquad
 g=\left(\begin{array}{cc}
 \alpha & \beta \\ \bar{\beta} & \bar{\alpha}
 \end{array}\right)\in SU(1,1).
 \eb
 Any gauge field configuration corresponding to a single vortex and a
 uniform magnetic field can be reduced to
 $\mathcal{A}=\mathcal{A}^{(B)}+\mathcal{A}^{(v)}$, using the
 transformation (\ref{su11}) combined with a gauge change.

 The Hamiltonian (\ref{hamgen}) in the presence of a vortex has
 thus the following form:
 \be\label{hamvort}
 \hat{H}_v=-\frac{\left(1-r^2\right)^2}{R^2}\left\{\partial_{rr}+\frac{1}{r}\,\partial_r
 +\frac{1}{r^2}\,(\partial_{\varphi}+i\nu)^2+\frac{4ib}{1-r^2}\,(\partial_{\varphi}+i\nu)-
 \frac{4b^2}{\left(1-r^2\right)^2}\,r^2\right\}.
 \eb
  This Hamiltonian still commutes with the angular momentum
 operator $\hat{L}$. Radial Hamiltonians $\hat{H}_{v,l}$ are obtained
 by the restriction of $\hat{H}_v$  to the
 eigenspaces of $\hat{L}$ with fixed angular momenta
 $l\in\Zb$. Namely, one obtains $\hat{H}_{v,l}=\hat{H}_{l+\nu}$, where
 the operators $\hat{H}_{\alpha\in\Rb}$ are defined as in
 (\ref{hamufml}). Thus the only effect the
 AB vortex has on the formal Hamiltonians is the shift
 of the angular momentum variable by $\nu$. This observation
 allows to considerably simplify the derivation of many results, using the
 calculations from the previous section.\vspace{0.1cm}\\
 \textbf{Remark}. As usual, for integer flux values some further simplifications occur.
 The Hamiltonians
 $\hat{H}^{(B)}$ and $\hat{H}_v$
 are related by a gauge transformation
 \be\label{gauge}
 \hat{H}_v=U\hat{H}^{(B)}U^{\dag},\qquad
 U:w\mapsto e^{-i\nu\varphi}w,
 \eb
 which is globally well-defined for $\nu\in\Zb$.
 The kernels of the resolvents of $\hat{H}^{(B)}$ and $\hat{H}_v$
 in this case differ only by a factor of $e^{i\nu(\varphi-\varphi')}$, and this change
 has no effect on the observable quantities.
 \subsection{Self-adjointness}
 From now on it will be assumed that $-1<\nu\leq0$ (it is clear
 from the above that this involves no loss of generality). Let us consider
 the full Hamiltonian $\hat{H}_v$ and restrict its domain to
 functions with compact support on the punctured disk:
 $\mathcal{D}(\hat{H}_v)=C_0^{\infty}(D\backslash\{0\})$. It was
 shown in the previous section that for $|l|\geq1$ the operator
 $\hat{H}_l$ is essentially self-adjoint, and for $|l|<1$ it has
 deficiency indices $(1,1)$. One should then distinguish two cases:
 \begin{itemize}
 \item \underline{$\nu=0$}. In this case $\hat{H}_v$ has
 deficiency indices $(1,1)$ and admits a one-parameter family of
 SAEs $\hat{H}^{(\gamma)}_v$ with $\gamma\in[0,2\pi)$ and
 \ben
 \mathcal{D}(\hat{H}^{(\gamma)}_v)=
 \left\{f+c\left(w_0^++e^{i\gamma }w_0^-\right)\,|\,
 f\in {C}_0^{\infty}(D\backslash\{0\}),c\in\Cb\right\}.
 \ebn
 These Hamiltonians describe a purely contact (non-magnetic)
 interaction of a particle with the AB solenoid.
 They have already been considered
 in \cite{aeg}, so we will not pursue their study.
 \item \underline{$-1<\nu<0$}. For such $\nu$ the deficiency subspaces $\mathcal{K}^{\pm}$ of
 the full Hamiltonian $\hat{H}_v$  are generated by those of the
 operators $\hat{H}_{\nu}$ and $\hat{H}_{1+\nu}$. Thus $\hat{H}_v$
 has deficiency indices $(2,2)$ and admits a four-parameter family
 of SAEs. Different extensions can be labeled by a unitary $2\times2$ matrix $U$
 and characterized by the domains
 \ben
 \mathcal{D}(\hat{H}^{\,U}_{\,v})=\left\{f+\sum_{i=1,2}
 c_i\Bigl(\mathrm{w}^+_i + \sum_{j=1,2}U_{ij}\,\mathrm{w}^-_j\Bigr)\,\Bigl|\Bigr.\,
 f\in {C}_0^{\infty}(D\backslash\{0\}),c_{1,2}\in\Cb\right\}.
 \ebn
 where $\mathrm{w}^{\pm}_{1,2}$ are orthonormal elements of the bases of
 $\mathcal{K}^{\pm}$,
 \ben
 \mathrm{w}^{\pm}_1(t,\varphi)=\frac{w^{\pm}_{\nu}(t)}{\|w^{\pm}_{\nu}(t)\|},\qquad
 \mathrm{w}^{\pm}_2(t,\varphi)=\frac{w^{\pm}_{1+\nu}(t)}{\|w^{\pm}_{1+\nu}(t)\|}\,e^{i\varphi},
 \ebn
 and $\|\cdot\|$ denotes the $L^2$-norm on $I$ with respect to the
 measure $d\mu_t$.
 \end{itemize}
 Note that the diagonal matrix $U$ describes magnetic point interactions acting
 separately in $s$-channel ($l=0$) and $p$-channel ($l=1$). Non-diagonal
 $U$ introduces a coupling between the two modes so that
 the Hamiltonian no longer commutes with the angular momentum.

 Further analysis of spectral properties of $H^{U}_v$ is a bit
 cumbersome in the general case (see, for example, the papers \cite{stovicek2},
 \cite{adami,stovicek1}, where such an analysis has been
 performed for the AB effect on the plane with
 and without magnetic field). We remark,
 however, that there exists a distinguished SAE of $\hat{H}_v$,
 whose domain consists of functions vanishing for $t\rightarrow0$.
 This extension will be  denoted by $\hat{H}_v^{\text{reg}}$.
 The next section is devoted to the calculation of its resolvent
 $\left(\hat{H}_v^{\text{reg}}-k^2\right)^{-1}$.  The resolvent of any other SAE
 can be obtained from the latter using Krein's formula \cite{albeverio}.

 \section{One-vortex resolvent}
 \subsection{Contour integral representations of the radial
 waves}
 The main technical difficulty in the calculation of the resolvent kernel
 $G_{k}(z,z')$ of the Hamiltonian $\hat{H}_v^{\text{reg}}$
 is the summation of  radial contributions coming from different angular momenta:
 \be\label{gfgen}
 G_k(z,z')=\frac{1}{2\pi}\sum\limits_{l\in\Zb}G_{k,l+\nu}(t,t')e^{il(\varphi-\varphi')}.
 \eb
 In order to address this problem, it is useful to introduce instead of the radial waves
 (\ref{sqis})--(\ref{ausol2}) the functions depending on both $t$
 and $\varphi$:
 \begin{eqnarray}
 \label{pw1}\mathrm{w}_l^{(I)}(z)&=&\frac{\Gamma(\chi+b)\Gamma(\chi-b)}{\Gamma
 (2\chi)}\,e^{\;il(\varphi+\pi)}\,w_l^{(I)}(t),\\
 \label{pw2}\hat{\mathrm{w}}_l^{(I)}(z)&=&\frac{\Gamma(\chi+b)\Gamma(\chi-b)}{\Gamma
 (2\chi)}\,e^{-il(\varphi+\pi)}w_l^{(I)}(t),
 \end{eqnarray}
 \begin{eqnarray}
 \label{pw3}\mathrm{w}_l^{(II,\pm)}(z)&=&2\pi i \,\frac{\Gamma(\chi\mp b\pm l)}{\Gamma
 (\chi\mp b)\Gamma(1\pm l)}\,e^{\;il(\varphi+\pi)}\,w_l^{(II,\pm)}(t),\\
 \label{pw4}\hat{\mathrm{w}}_l^{(II,\pm)}(z)&=&2\pi i\, \frac{\Gamma(\chi\mp b\pm l)}{\Gamma
 (\chi\mp b)\Gamma(1\pm l)}\,e^{-il(\varphi+\pi)}w_l^{(II,\pm)}(t).
 \end{eqnarray}
 Combining these formulas with the relations
 (\ref{gfans}), (\ref{wronsk121}), one can rewrite the Green
 function (\ref{gfgen}) in the following way:
 \be\label{intergf}
 G_k(z,z')=\frac{e^{-i\nu(\varphi-\varphi')}}{8i\pi^2}\left(\mathcal{G}^{(+)}_k(z,z')+
 \mathcal{G}^{(-)}_k(z,z')\right),
 \eb
 where the functions $\mathcal{G}^{(\pm)}_k(z,z')$ are given by
 \begin{eqnarray}
 \label{gpm1}
 \mathcal{G}^{(\pm)}_k(z,z')&=&
 \sum\limits_{{l\in\Zb+\nu,\; l\gtreqless0}}
 \mathrm{w}_l^{(I)}(z)\hat{\mathrm{w}}_l^{(II,\pm)}(z')\qquad \text{for }|z|>|z'|,\\
 \label{gpm2}
 \mathcal{G}^{(\pm)}_k(z,z')&=&
 \sum\limits_{{l\in\Zb+\nu,\; l\gtreqless0}}
 \mathrm{w}_l^{(II,\pm)}(z)\hat{\mathrm{w}}_l^{(I)}(z')\qquad \text{for }|z|<|z'|.
 \end{eqnarray}

 The sums (\ref{gpm1})--(\ref{gpm2}) can be computed using a special set of solutions
 of stationary Schr\"odinger
 equation without AB flux, known as horocyclic waves \cite{DeMicheli}.
 These solutions have the form
 \be\label{horowaves}
 \Psi_{\pm}(z,\theta)=\frac{\left(1-|z|^2\right)^{\chi_{\pm}}}{\left(1+z\,e^{-\theta}\right)^{\chi_{\pm}-b}
 \left(1+\bz\, e^{\theta}\right)^{\chi_{\pm}+b}}\,,
 \eb
 where
 \ben
 \chi_{\pm}=\frac12\pm\left(\chi-\frac12\right)
 \ebn
 and $\theta$ is an arbitrary complex parameter. Being considered as functions of $\theta$,
 horocyclic waves $\Psi_{\pm}(z,\theta)$ have an infinite number of branchpoints located at $\theta=\pm\ln
 r+i\left(\varphi+\pi+2\pi\Zb\right)$.
 Let us introduce a
 system of branch cuts in the $\theta$-plane as shown in the
 Fig.~1. The sheets of Riemann surfaces of the
 functions $\Psi_{\pm}(z,\theta)$ are fixed by the requirement that the
 arguments of both $1+z\,e^{-\theta}$ and $1+\bz\,e^{\theta}$ are
 equal to zero on the line $\mathrm{Im}\,\theta=\varphi$.
 \begin{figure}[h]
 \begin{center}
 \resizebox{9cm}{!}{
 \includegraphics{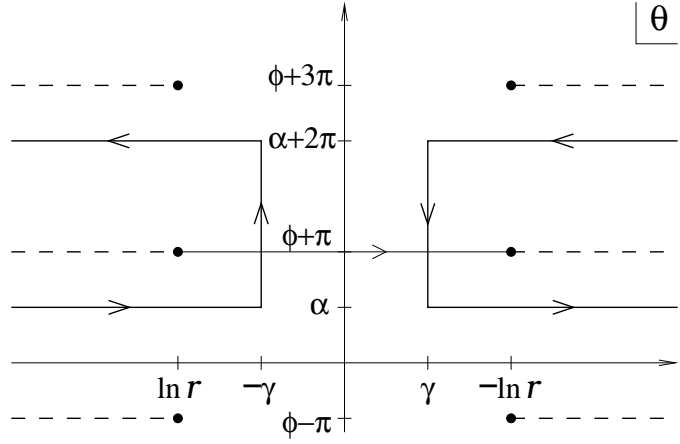}}\vspace{0.2cm}\\
 \end{center}
 \caption{Contours of integration in the $\theta$-plane}
 \end{figure}

 Recall that the Hamiltonians $\hat{H}^{(B)}$ and $\hat{H}_v$ are
 related by the gauge transformation (\ref{gauge}). Although this
 transformation is singular for non-integer values of the flux,
 one can still relate any solution of the equation $(\hat{H}_v-k^2)w=0$
 to a solution of the same equation without AB
 field, $(\hat{H}^{(B)}-k^2)\psi=0$. However, since we have
 $w=e^{-i\nu\varphi}\psi$, the function $\psi$ should be branched with the monodromy $e^{2\pi i \nu}$ at the
 point $z=0$. Motivated by this well-known fact, we will try to
 represent radial wave functions (\ref{pw1})--(\ref{pw4}) as superpositions of
 elementary solutions (\ref{horowaves}),
 \ben
 \mathrm{w}(z)=\int\nolimits_C\Psi_{\pm}(z,\theta)\,\rho(\theta)\,d\theta\,,
 \ebn
 where $C$ is an integration contour and $\rho(\theta)$ is an appropriately chosen weight function.
 There will be three types of contours that will be important to us (see also Fig.~1):
 \begin{itemize}
 \item Contour $C_+(z)$ starts at $-\infty+i\alpha$,
 surrounds the branch cut
 $\mathbf{b}_+=\bigl(-\infty+i(\varphi+\pi),\ln r+i(\varphi+\pi)\bigr]$
 in a counter-clockwise manner, and goes to $-\infty+i(\alpha+2\pi)$.
 \item Contour $C_-(z)$  starts at $\infty+i(\alpha+2\pi)$, then
 goes counter-clockwise around the branch cut
 $\mathbf{b}_-=\bigl[-\ln
 r+i(\varphi+\pi),\infty+i(\varphi+\pi)\bigr)$, and finally
 travels to $\infty+i\alpha$ along the ray parallel to the real
 axis.
 \item Contour $C_0(z)$ joins two
 branchpoints:
 $\theta_1=\ln r+i(\varphi+\pi)$ and $\theta_2=-\ln r+i(\varphi+\pi)$.
 \end{itemize}
 Real parameters $\alpha$ and $\gamma$ can be chosen
 arbitrarily; the only conditions they should satisfy are given by
 \ben
 |\varphi-\alpha|<\pi,\qquad 
 0\leq\gamma<-\ln r.
 \ebn

 Assuming that $\mathrm{Re}\,k^2<0$, one may now write a number of contour integral representations
 for the radial waves (\ref{pw1})--(\ref{pw4}):
 \begin{eqnarray}
 \label{int1}\mathrm{w}^{(I)}_l(z)&=&\int\nolimits_{C_0(z)}\Psi_-(z,\theta)\,e^{l\theta}\,d\theta,\\
 \label{int2}\hat{\mathrm{w}}^{(I)}_l(z)&=&\int\nolimits_{C_0(z)}\hat{\Psi}_-(z,\theta)\,e^{-l\theta}\,d\theta,\\
 \label{int3}\mathrm{w}^{(II,\pm)}_l(z)&=&\pm\int\nolimits_{C_{\pm}(z)}\Psi_+(z,\theta)\,e^{l\theta}\,d\theta,\\
 \label{int4}\hat{\mathrm{w}}^{(II,\pm)}_l(z)&=&\mp\int\nolimits_{C_{\mp}(z)}\hat{\Psi}_+(z,\theta)\,
 e^{-l\theta}\,d\theta,
 \end{eqnarray}
 where the functions $\hat{\Psi}_{\pm}(z,\theta)$ are obtained
 from ${\Psi}_{\pm}(z,\theta)$ by replacing $b\rightarrow -b$.
 Although the validity of the representations (\ref{int1})--(\ref{int4}) can be checked
 directly, their general structure may also be a posteriori
 understood as follows. Consider, for instance, the functions $\mathrm{w}^{(I)}_l(z)$ and
 $\mathrm{w}^{(II,\pm)}_l(z)$ as defined by (\ref{int1}) and (\ref{int3}). Continuation
 of these functions along a counter-clockwise circuit
 enclosing the point $z=0$ amounts to simultaneous shift of the branch cuts and integration contours
  upwards by $2\pi$ in the $\theta$-plane. This shift is in turn equivalent to simple multiplication
 of both functions by $e^{2\pi i l}$. Moreover, elementary solutions (\ref{horowaves})
 satisfy the relation
 \ben
 \hat{L}\Psi_{\pm}(z,\theta)=(z\partial_z-\bz\partial_{\bz})\Psi_{\pm}(z,\theta)
 =-\partial_{\theta}\Psi_{\pm}(z,\theta),
 \ebn
 which means that RHSs of (\ref{int1}) and (\ref{int3}) are common
 (multivalued) eigenfunctions of $\hat{H}^{(B)}$ and $\hat{L}$, their angular
 momenta being equal to $l$. First function is regular for
 $t\rightarrow1$, since in this case the branch cuts pinch the
 imaginary axis. Similarly, the second function is regular for
 $t\rightarrow0$. This implies (modulo
 constant factors that have to be found by a direct calculation)
 the relations (\ref{pw1}) and (\ref{pw3}).
 \subsection{Summation}
 Let us now turn to the calculation of the sums
 (\ref{gpm1})--(\ref{gpm2}). For simplicity the case $|z|>|z'|$ is
 treated in detail and we only indicate the changes needed to
 handle another case. Substituting contour representations
 (\ref{pw1}) and (\ref{pw4}) into the relation (\ref{gpm1}), one
 obtains
 \ben
 \mathcal{G}^{(\pm)}_k(z,z')=\mp\sum\limits_{l\in\Zb+\nu,\;l\gtreqless0}\int\nolimits_{C_0(z)}\!\!\!d\theta_1
 \int\nolimits_{C_\mp(z')}\!\!\!d\theta_2
 \;\Psi_-(z,\theta_1)\hat{\Psi}_+(z',\theta_2)\;e^{l(\theta_1-\theta_2)}.
 \ebn
 Since $|z|>|z'|$, one may choose the contours $C_{\pm}(z')$ in
 such a way that $\gamma_{z'}>-\ln r$. Consequently, we have
 $\mathrm{Re}\left(\theta_1-\theta_2\right)<0$ for all $\theta_1\in
 C_0(z)$, $\theta_2\in C_-(z')$ and $\mathrm{Re}\left(\theta_1-\theta_2\right)>0$ for all $\theta_1\in
 C_0(z)$, $\theta_2\in C_+(z')$. Then it becomes possible to
 perform the summation inside the integrals and one finds
 \ben
 \mathcal{G}^{(+)}_k(z,z')+\mathcal{G}^{(-)}_k(z,z')=
 \int\limits_{C_0(z)}\!\!d\theta_1\!\!\!\!
 \int\limits_{C_+(z')\cup C_-(z')}\!\!\!\!\!\!\!\! d\theta_2
 \;\;\;\Psi_-(z,\theta_1)\hat{\Psi}_+(z',\theta_2)\;
 \frac{e^{(1+\nu)(\theta_1-\theta_2)}}{e^{\theta_1-\theta_2}-1}
 \ebn
 We would like to deform the contours $C_{\pm}(z')$ in the last
 integral over $\theta_2$ so that their vertical parts compensate
 one another. Then $C_+(z')\cup C_-(z')$ transforms into two
 horizontal lines, but one also earns a pole contribution coming from
 $e^{\,\theta_2}=e^{\,\theta_1}$. Next, if we assume that $\varphi-\varphi'\neq\pm\pi$,
 then the two lines can be deformed into $\mathrm{Im}\,\theta_2=\varphi'$
 using quasiperiodicity in $\theta_2$.
 Together with (\ref{intergf}), this leads to the following representation for
 the Green function:
 \be\label{ans}
 G_k(z,z')=\left\{
 \begin{array}{rl}
 e^{- i \nu(\varphi-\varphi'+2\pi)}\,G_k^{(0)}(z,z')+\Delta_k(z,z') & \text{for }\varphi-\varphi'\in(-2\pi,-\pi),\\
 e^{- i \nu(\varphi-\varphi')}\; G_k^{(0)}(z,z')+\Delta_k(z,z') & \text{for }\varphi-\varphi'\in(-\pi,\pi),\\
 e^{- i \nu(\varphi-\varphi'-2\pi)}\,G_k^{(0)}(z,z')+\Delta_k(z,z') & \text{for
 }\varphi-\varphi'\in(\pi,2\pi)\,,
 \end{array}\right.
 \eb
 with
 \be\label{g01}
 G_k^{(0)}(z,z')=\frac{1}{4\pi}\int\nolimits_{C_0(z)}\!\!\!d\theta\;\;
 \Psi_-(z,\theta)\hat{\Psi}_+(z',\theta)\,,
 \eb
 \be\label{delta1}
 \Delta_k(z,z')=\frac{1-e^{-2\pi i\nu}}{8i\pi^2}\;e^{- i \nu(\varphi-\varphi')}
 \int\limits_{C_0(z)}\!\!d\theta_1\!\!\!\!
 \int\limits_{\mathrm{Im}\,\theta_2=\varphi'}\!\!\!\!\! d\theta_2
 \;\;\Psi_-(z,\theta_1)\hat{\Psi}_+(z',\theta_2)\;
 \frac{e^{(1+\nu)(\theta_1-\theta_2)}}{e^{\,\theta_1-\theta_2}-1}\,.
 \eb

 Similarly, assuming that $|z|<|z'|$, one obtains an integral representation
 of the Green function which has exactly the same form as
 (\ref{ans}), except that the functions $G_k^{(0)}(z,z')$ and
 $\Delta_k(z,z')$ are now given by
  \be\label{g02}
 G_k^{(0)}(z,z')=\frac{1}{4\pi}\int\nolimits_{C_0(z')}\!\!\!d\theta\;\;
 \hat{\Psi}_-(z',\theta)\Psi_+(z,\theta)\,,
 \eb
 \be\label{delta2}
 \Delta_k(z,z')=\frac{e^{2\pi i\nu}-1}{8i\pi^2}\;e^{- i \nu(\varphi-\varphi')}
 \int\limits_{C_0(z')}\!\!d\theta_1\!\!\!\!
 \int\limits_{\mathrm{Im}\,\theta_2=\varphi}\!\!\!\!\! d\theta_2
 \;\;\hat{\Psi}_-(z',\theta_1)\Psi_+(z,\theta_2)\;
 \frac{e^{(1+\nu)(\theta_2-\theta_1)}}{e^{\,\theta_2-\theta_1}-1}\,.
 \eb

 After some computations (technical details are
 outlined in the Appendix~A) one may show that both representations
 coincide. Moreover, the integrals (\ref{g01}) and (\ref{g02}) can be carried out
 explicitly:
 \be
 \label{g0ans}
 G_k^{(0)}(z,z')=
 \left(\frac{1-z\bz'}{1-\bz z'}\right)^b\zeta\Bigl(u(z,z')\Bigr),
 \eb
 where $\displaystyle u(z,z')=\left|\frac{z'-z}{1-\bz z'}\right|^2 $
 has a simple relation with the geodesic distance between the points $z$ and
 $z'$, and the function $\zeta(u)$ is given by
 \be\label{g0anss}
 \zeta(u)=\frac{1}{4\pi}\frac{\Gamma(\chi+b)\Gamma(\chi-b)}{\Gamma(2\chi)}\,
 \Bigl(1-u\Bigr)^{\chi}\;{}_2F_1\Bigl(\chi+b,\chi-b,2\chi,1-u\Bigr).
 \eb
 Note that $G_k^{(0)}(z,z')$ coincides with the well-known expression for the
 resolvent kernel of the Hamiltonian without AB field
 \cite{comtet,elstrodt}. This can also be seen directly from the
 representation (\ref{ans}), since $\Delta_k(z,z')$ in (\ref{delta1}) or (\ref{delta2})
 obviously vanishes for $\nu=0$.
 The function $\Delta_k(z,z')$
 may also be written in a symmetric form:
 \be\label{deltaans}
 \Delta_k(z,z')=\frac{\sin\pi\nu}{\pi}\int\limits_{-\infty}^{\infty}d\theta\;
 \frac{e^{(1+\nu)\theta+i(\varphi-\varphi')}}{1+e^{\,\theta+i(\varphi-\varphi')}}\,
 \left(\frac{1+rr'e^{-\theta}}{1+rr'e^{\;\theta}\;}\right)^b\zeta
 \Bigl(v(r,r',\theta)\Bigr),
 \eb
 with
 \be\label{vrr}
 v(r,r',\theta)=\frac{r^2+r'^2+2rr'\cosh\theta}{1+r^2r'^2+2rr'\cosh\theta}\,.
 \eb
 In our opinion, the representation (\ref{ans}) and the formulas
 (\ref{g0ans})--(\ref{vrr}) constitute the most interesting results of the
 present paper. It is instructive to compare them with the known results in
 the flat space (cf. the relations (2.25)--(2.26) in \cite{pacific} or
 the formula (5.10) from \cite{marino}). Notice that  the `free' part
 of the Green function is manifestly separated in (\ref{ans})
 from the vortex-dependent contribution
 $\Delta_k(z,z')$.

 \section{Spectrum and density of states}
 The spectrum of the regular extension $\hat{H}_v^{\text{reg}}$
 consists of three parts:
 \begin{itemize}
 \item a continuous spectrum $\displaystyle
 E\in\left[{(1+4b^2)}/{R^2}\,,\infty\right)$;
 \item
 a finite number of infinitely degenerate eigenvalues, which
 coincide with the usual Landau levels on the hyperbolic disk
 \cite{comtet,grosche} in the absence of the AB field.
 These levels are explicitly given by
 \be\label{energy1}
 E_n^{(0)}=\frac{1}{R^2}\left[1+4b^2-4\left(|b|-n-\frac{1}{2}\right)^2\right],
 \eb
 where $n=0,1,\ldots , n_{max}<|b|-1/2$. Corresponding common
 eigenfunctions of the Hamiltonian $\hat{H}_v^{\text{reg}}$ and the
 angular momentum operator $\hat{L}$ can be expressed in terms of Jacobi's
 polynomials (cf. the relation (13) in \cite{grosche}):
 \ben
 \Psi_{n,l}^{(0)}(t,\varphi)\sim t^{|l+\nu|/2}(1-t)^{|b|-n}P_n^{\left(2|b|-2n-1,|l+\nu|\right)}(2t-1)\;e^{il\varphi}.
 \ebn
 Here one should take $l=0,-1,-2,\ldots$ for $b>0$ and $l=1,2,\ldots$ for
 $b<0$.
 \item a finite number of bound
 states $E_n^{(\nu)}$ with finite degeneracy. The form of these eigenvalues depends
 on the sign of magnetic field. Namely, for $b>0$ one has
 \be\label{energy2}
 E_n^{(\nu,+)}=\frac{1}{R^2}\left[1+4b^2-4\left(b-n-(1+\nu)-\frac{1}{2}\right)^2\right],
 \eb
 where $n=0,1,\ldots,n'_{max}<b-(\nu+1)-1/2$. In the case $b<0$,
 the eigenvalues may be written as
 \be\label{energy3}
 E_n^{(\nu,-)}=\frac{1}{R^2}\left[1+4b^2-4\left(|b|-n+\nu-\frac{1}{2}\right)^2\right],
 \eb
 with $n=0,1,\ldots,n''_{max}<|b|+\nu-1/2$. Common eigenstates of $\hat{H}_v^{\text{reg}}$
 and $\hat{L}$ are again given by Jacobi's polynomials:
 \begin{eqnarray*}
  b>0:\quad\Psi_{n,l}^{(\nu,+)}(t,\varphi)&\sim& t^{(l+\nu)/2}(1-t)^{b-n-(\nu+1)}
  P_n^{\left(2b-2n-2(\nu+1)-1,l+\nu\right)}(2t-1)\;e^{il\varphi},\\
 b<0:\quad\Psi_{n,l}^{(\nu,-)}(t,\varphi)&\sim& t^{|l+\nu|/2}\,(1-t)^{|b|-n+\nu}
 \quad
 \,P_n^{\left(2|b|-2n+2\nu-1,|l+\nu|\right)}(2t-1)\;e^{il\varphi}.
 \end{eqnarray*}
 For given radial quantum number $n$ the allowed eigenvalues of the angular
 momentum are $l=1,2,\ldots,n+1$ (for $b>0$) and $l=0,-1,\ldots,-n$
 (for $b<0$).
 \end{itemize}
 \textbf{Remark}. The above expressions (\ref{energy1})--(\ref{energy3}) for the energy levels
 can also be extracted from the recent work \cite{bulaev}. It is worthwhile to
 emphasize that the discrete spectrum is absent for $|b|<{1}/{2}$.\vspace{0.1cm}

 Let us now consider the density of states (DoS) on the hyperbolic disk.
 It can  be obtained from the boundary values of the resolvent kernel
 on the real axis in the complex energy plane, using the following
 formula:
 \ben
 \rho(E)=\frac{1}{\pi}\;\mathrm{Im}\,\mathrm{Tr}\,G_k(z,z'\rightarrow
 z)\Bigl|_{\;k^2=E+i0}\Bigr. ,\qquad E\in\Rb.
 \ebn

 Both terms in the
 representation (\ref{ans}) of the Green function contribute to
 the DoS.
 The contribution of the free-resolvent kernel $G^{(0)}_k(z,z')$ has been first calculated
 by Comtet \cite{comtet}. His results (supplemented by an
 additional term  \cite{bgm}, coming from the discrete spectrum)
 give the following expression for the DoS:
 \begin{eqnarray*}
 \rho^{(0)}(E,z)&=&\frac{1}{\pi}\;\mathrm{Im}\,G^{(0)}_k(z,z'\rightarrow
 z)\Bigl|_{\;k^2=E+i0}\Bigr.\,=\\
 &=&\frac{1}{4\pi}\frac{\sinh 2\pi\lambda}{\cosh2\pi\lambda+\cos 2\pi
 b}\;\Theta\left(E-\frac{1+4b^2}{R^2}\right)+\\
 &\;&+\;\frac{2}{\pi R^2}\sum_{n=0 }^{n_{max}}
 \left(|b|-n-\frac12\right)\delta\left(E-E^{(0)}_n\right).
 \end{eqnarray*}
 Here $\Theta(x)$ denotes Heaviside function and
 \be\label{lambda}
 \lambda=\frac12\sqrt{ER^2-1-4b^2}\,.
 \eb

 One can not expect that the DoS per unit area, induced by the
 AB field, will also be constant on $D$. However, it should depend only on the geodesic
 distance between a given point on the disk and the flux position. Indeed, since the function $\Delta(z,z')$
 is non-singular for $z\rightarrow z'$, the vortex-dependent part of the DoS
 is given by
 \be\label{auxi1}
 \rho^{(\nu)}(E,z)=\frac{1}{\pi}\,\mathrm{Im}\,\Delta_k(t)\Bigl|_{\;k^2=E+i0}\Bigr.\,,
 \eb
 where the function $\Delta_k(t)$ is obtained from $\Delta(z,z')$
 by setting $\varphi=\varphi'$,
 $r^2=r'^2=t$:
 \be\label{deltaansred}
 \Delta_k(t)=\frac{\sin\pi\nu}{\pi}\int\limits_{-\infty}^{\infty}d\theta\;
 \frac{\;e^{\,(1+\nu)\theta}}{1+e^{\,\theta}}\,
 \left(\frac{1+te^{-\theta}}{1+te^{\;\theta}\;}\right)^b\zeta
 \left(\frac{2t(1+\cosh\theta)}{1+t^2+2t \cosh \theta}\right).
 \eb
 As it stands, the representation (\ref{deltaansred}) is valid
 in the left half-plane  $\mathrm{Re}\,k^2<0$, where the function  $\Delta_k(t)$ is analytic.
 However, the DoS is determined by the singularities of
 $\Delta_k(t)$ that occur on the positive part of the real axis
 (we may expect there a
 finite number of poles and
 the branch cut $\bigl[{(1+4b^2)}/{R^2}\,,\infty\bigr)$,
 corresponding to the continuous part of the spectrum
 of $\hat{H}_v^{\text{reg}}$). One could try to construct the
 appropriate
 analytic continuation of  $\Delta_k(t)$, considering (\ref{deltaansred}) as a
 contour integral and then suitably deforming the
 contour. It seems, however, that this approach
 does not lead to any satisfactory result because of the complicated
 singularity structure  of the function
 under the integral sign in the $\theta$-plane.

 An alternative method consists in the following. Remark
 that the vortex-dependent contribution to the DoS in the \underline{whole} hyperbolic space
 \be\label{auxi2}
 \rho^{(\nu)}(E)=\int\nolimits_{D}d\mu \;\rho^{(\nu)}(E,z)
 \eb
 has a finite value, since
 \begin{eqnarray*}
 &\;&\left(\frac{1+te^{-\theta}}{1+te^{\;\theta}\;}\right)^b\zeta
 \left(\frac{2t(1+\cosh\theta)}{1+t^2+2t \cosh
 \theta}\right)=\\ &=&\left\{\begin{array}{l}
 \displaystyle -\frac{1}{4\pi}\Bigl[\ln 2t+\ln(1+\cosh\theta) +2\gamma_{E}+\psi(\chi+b)+\psi(\chi-b)\Bigr]+o(1)\;
 \text{ for } t\rightarrow0,\\ \\
 \displaystyle\frac{1}{4\pi}\frac{\Gamma(\chi+b)\Gamma(\chi-b)}{\Gamma(2\chi)}
 \frac{(1-t)^{2\chi}}{(1+e^{\theta})^{\chi+b}(1+e^{-\theta})^{\chi-b}}+
 o\left((1-t)^{2\chi}\right)\;\qquad\text{ for } t\rightarrow 1.
 \end{array}\right.
 \end{eqnarray*}
 If one now integrates $\Delta_k(t)$ over spatial
 coordinates (see Appendix~B) and then considers the analytic continuation of the
 result to the complex energy plane, the following expression
 for $\rho^{(\nu)}(E)$ can be obtained:
  \begin{eqnarray}\label{finresd}
  \rho^{(\nu)}(E)=-\frac{R^2}{4\pi}\;\mathrm{Im}\,\frac{1}{2\chi-1}
 \left\{(\chi-b+\nu)\Bigl[\psi(\chi-b)-\psi(\chi-b+\nu+1)\Bigr]+\right.&\;&\\
 \nonumber \left. +\;
 (\chi+b-\nu-1)\Bigl[\psi(\chi+b)-\psi(\chi+b-\nu-1)\Bigr]\right\}\Bigl|_{k^{2}=E+i0}&=&\\
 \nonumber =\rho^{(\nu)}_d(E)+\rho^{(\nu)}_c(E),
 \end{eqnarray}
 where the contributions of the discrete and continuous part of
 the spectrum are given by
 \be\label{rhopoint}
 \rho^{(\nu)}_d(E) =\left\{\begin{array}{l}
 \displaystyle\sum_{n=0}^{\;n_{max}'}(n+1)\,\delta\left(E-E^{(\nu,+)}_n\right)-
 \sum_{n=0}^{\;n_{max}}(n-\nu)\,\delta\left(E-E^{(0)}_n\right)\quad\;\;\,\qquad \text{for }
 b>0,\\
  \displaystyle\sum_{n=0}^{\;n_{max}''}(n+1)\,\delta\left(E-E^{(\nu,-)}_n\right)-
 \sum_{n=0}^{\;n_{max}}(n+\nu+1)\,\delta\left(E-E^{(0)}_n\right)\qquad \text{for }
 b<0,
 \end{array}\right.
 \eb
 \begin{eqnarray}
 \label{rhocont}\rho^{(\nu)}_c(E)=-\frac{R^2}{8\lambda}\;\Theta\left(E-\frac{1+4b^2}{R^2}\right)
 \left\{\frac{\lambda\sinh2\pi\lambda+\left(\frac12-b+\nu\right)\sin2\pi(b-\nu)}{\cosh2\pi\lambda+\cos2\pi (b-\nu)}\,
 -\;\;\right.\\
 \nonumber-\left.\frac{\lambda\sinh2\pi\lambda+\left(\frac12-b+\nu\right)\sin2\pi b}{\cosh2\pi\lambda+\cos2\pi b}
 \right\},
 \end{eqnarray}
 and the parameter $\lambda$ is defined as in (\ref{lambda}).

 At last we add a comment concerning the flat space limit
 ($R\rightarrow\infty$)
 at zero magnetic field ($b=0$). In this case the representation (\ref{finresd}) for the
 vortex-dependent DoS transforms into
 \be\label{flsp}
  \rho^{(\nu)}(E)\;\substack{R\rightarrow\infty \\ \longrightarrow \\
  \;}\;\frac{1}{\pi}\;\mathrm{Im}\,\frac{\nu(\nu+1)}{2k^2}\Bigl|_{k^2=E+i0}\Bigr.=-\frac{\nu(\nu+1)}{2}\,\delta(E)\,.
 \eb
 This result has been first obtained in \cite{comtet_jpa}, and it has important consequences in the
 theory of disordered magnetic systems (see, for example, \cite{desbois1,desbois2,desbois3}). Obtaining
 the relation (\ref{flsp}) directly from (\ref{rhocont}) is
 more subtle; one should consider $\rho^{(\nu)}_c(E)$
 as a distribution and  supply it with a proper regularization
 at the edge of the spectrum, i.~e. as $\lambda\rightarrow+0$.

 \section{Discussion}
 We have studied the Hamiltonian of a particle moving on
 the hyperbolic disk in the background of a uniform magnetic field
 and the AB gauge potential. The
 density of states and the resolvent of this operator have been calculated
 in a closed form, using Sommerfeld-type
 integral representations for the radial waves.

 The above discusssion does not exhaust all problems related to
 the system under consideration.
 First of all, there remains a technical question of analytic
 continuation of the representation (\ref{deltaans}) to energy values with
 $\mathrm{Re}\,k^2\geq0$. Such continuation would allow to
 investigate the curvature dependence of various vacuum quantum numbers,
 e.~g.
 fractional charge, magnetic flux and angular momentum (see, for
 instance, \cite{sitenko}), induced by the AB vortex. One may also
 try to address the latter problem, applying the technique
 we have used in the calculation of the density of
 states.

 Another interesting question concerns the generalization of our results to the
 case of the Dirac Hamiltonian. This problem, in its turn, appears to be non-trivially related to
 the theory of isomonodromic deformations
 on the hyperbolic disk \cite{narayanan,beatty,pt}. More precisely, it was shown in
 \cite{beatty} that the two-point isomonodromic tau function of the Dirac
 operator on the Poincar\'e disk provides a solution to
 Painlev\'e~VI equation. The explicit form of this Painlev\'e transcendent in a
 particular case of zero magnetic field has been conjectured by Doyon
 \cite{doyon}, whose idea was to replace the tau function by its physical analog ---
 a correlator of monodromy fields, and to sum up the corresponding form factor expansion.
 The missing ingredient in the rigorous proof of
 this result and its generalization to the case of non-zero field
 is the Green function of the Dirac operator with one branch point,
 precisely analogous to the resolvent found in the present paper.
 Moreover, the knowledge of this Green function enables one to compute all form
 factors of monodromy fields on the hyperbolic disk \cite{doyon},
 in particular, those of Ising spin and disorder fields \cite{doyon2,doyon_fonseca}.
 We leave these problems to a future publication.\vspace{0.2cm}\\
 \textbf{Acknowledgements}. I warmly thank Dublin Institute for Advanced Studies, where
 a large part of this work has been done, for hospitality. I am very grateful to Sasha
 Povolotsky for useful conversations and help with the literature.

\renewcommand{\theequation}{A.\arabic{equation}}
\setcounter{equation}{0}

 \appendix
 \section*{Appendix A}
 Here we describe a method of obtaining the expressions
 (\ref{g0ans})--(\ref{vrr}) for the Green function from the integral representations
 (\ref{g01})--(\ref{delta2}). First,
 note that horocyclic waves (\ref{horowaves}) satisfy the following
 identity:
 \be\label{horotr}
 \Psi_{\pm}(z,\theta)=
 {\left(\bar{\alpha}+\beta e^{-\theta}\right)^{-\chi_{\pm}+b}
 \left(\alpha+\bar{\beta}e^{\theta}\right)^{-\chi_{\pm}-b}}
 \left(\frac{\beta\bar{s}+\alpha}{\bar{\beta}s+\bar{\alpha}}\right)^b
 \Psi_{\pm}(s,\xi),
 \eb
 where we have introduced the notation
 \be\label{chva}
 s=\frac{\;\;\bar{\alpha}z-\beta}{-\bar{\beta}z+\alpha},\qquad
 e^{\,\xi}=\frac{\bar{\alpha}\,e^{\theta}+\beta}{\bar{\beta}\,e^{\theta}+\alpha}
 \eb
 and assumed that $\displaystyle |\alpha|^2+|\beta|^2=1$. Let us
 now make the change of variables (\ref{chva}) in (\ref{g01})
 (introducing similarly the variable
 $\displaystyle s'=\frac{\;\;\bar{\alpha}z'-\beta}{-\bar{\beta}z'+\alpha}$
 instead of $z'$). It is then straightforward to check that
 (\ref{g01}) reduces to
 \be\label{interm1}
 G_k^{(0)}(z,z')=\frac{1}{4\pi}
 \left(\frac{\beta\bar{s}+\alpha}{\bar{\beta}s+\bar{\alpha}} \right)^b
 \left(\frac{\bar{\beta}s'+\bar{\alpha}}{\beta\bar{s}'+\alpha} \right)^b
 \int\nolimits_{C_0(s)}\!\!\!d\xi\;\;
 \Psi_-(s,\xi)\hat{\Psi}_+(s',\xi)
 \eb
 We may now adjust the parameters $\alpha$ and $\beta$ in such a
 way that $s'=0$. For example, one can take
 \ben
 \alpha=\frac{1}{\sqrt{1-|z'|^2}}\,,\qquad  \beta=\frac{z'}{\sqrt{1-|z'|^2}}\,.
 \ebn
 Since in this case $\displaystyle s=\frac{z-z'}{1-z\bz'}$, the
 relation (\ref{interm1}) transforms into
 \begin{eqnarray}
 \nonumber G_k^{(0)}(z,z')=\frac{1}{4\pi}\left(\frac{1-z\bz'}{1-\bz
 z'}\right)^b\int\nolimits_{C_0(s)}\!\!\!d\xi\;\;
 \Psi_-(s,\xi) =\\
 \label{interm2} =\frac{1}{4\pi}\left(\frac{1-z\bz'}{1-\bz
 z'}\right)^b\left(1-|s|^2\right)^{\chi_-}\int\limits_{\ln|s|}^{-\ln|s|}
 \frac{d\xi}{\left(1-|s|e^{-\xi}\right)^{\chi_--b}\left(1-|s|e^{\,\xi}\right)^{\chi_-+b}}\,.
 \end{eqnarray}
 After the change of variable $\displaystyle
 e^{\,\xi}=\frac{1-(1-|s|^2)t}{|s|}$ in the last integral, we
 obtain standard representation of Gauss hypergeometric function,
 \ben
 G_k^{(0)}(z,z')=\frac{1}{4\pi}\left(\frac{1-z\bz'}{1-\bz
 z'}\right)^b\left(1-|s|^2\right)^{\chi}\int\limits_{0}^1\frac{t^{\chi-b-1}
 (1-t)^{\chi+b-1}dt}{\left(1-\left(1-|s|^2\right)t\right)^{\chi+b}}\,,
 \ebn
 which immediately gives (\ref{g0ans})--(\ref{g0anss}). Performing analogous manipulations
 with the representation
 (\ref{g02}), one finds the same answer. 


 Let us now consider the representation (\ref{delta1}) for the function $\Delta_k(z,z')$.
 Interchanging the order of integration and then introducing
 instead of  $\theta_2$ a new variable
 $\tilde{\theta}_2=\theta_2-\theta_1$, one can check that this relation transforms
 into
 \be\label{need1}
 \Delta_k(z,z')=\frac{1-e^{-2\pi i\nu}}{2\pi i}\; e^{-i\nu(\varphi-\varphi')}
 \!\!\!\!\!
 \int\limits_{\mathrm{Im}\,\tilde{\theta}_2=\varphi'-\varphi-\pi}
 \!\!\!\!\!d\tilde{\theta}_2
 \;\frac{e^{-(1+\nu)\tilde{\theta}_2}}{e^{-\tilde{\theta}_2}-1}\;F(z,z',\tilde{\theta}_2)\,,
 \eb
 where the function $F(z,z',\tilde{\theta}_2)$ after additional change of integration variable
 $\theta_1\rightarrow \tilde{\theta}_1=\theta_1+\tilde{\theta}_2$
 can be written as
 \be\label{intdel1}
 F(z,z',\tilde{\theta}_2)=\frac{1}{4\pi}\int\nolimits_{C_{\tilde{\theta}_2}(z)}\!\!\!
 d\tilde{\theta}_1\;\Psi_-(z,\tilde{\theta}_1-\tilde{\theta}_2)\hat{\Psi}_+(z',\tilde{\theta}_1)\,.
 \eb
 In the last expression, $C_{\tilde{\theta}_2}(z)$ represents the
 integration contour obtained from $C_{0}(z)$ by shifting it by $\tilde{\theta}_2$.

 One may now use the trick described above to eliminate the function
 $\hat{\Psi}_+(z',\tilde{\theta}_1)$ from (\ref{intdel1}). Namely,
 consider further change of variables:
 \ben
 \tilde{\theta}_1\rightarrow \xi,\qquad e^{\,\xi}=
 \frac{e^{\,\tilde{\theta}_1}+z'}{\bar{z}'e^{\,\tilde{\theta}_1}+1}\,.
 \ebn
 Somewhat tedious but fairly routine calculation
 shows that the integral (\ref{intdel1}), being rewritten in terms of $\xi$,
 reduces to
 \be\label{interm3}
 F(z,z',\tilde{\theta}_2)=\frac{1}{4\pi}\left(\frac{1+z'\tilde{s}}{1+\bz's}\right)^b
 \left(1-s\tilde{s}\right)^{\chi_-}\int\nolimits_{C_{\xi}}\frac{d\xi}{
 (1+s\,e^{-\xi})^{\chi_--b}(1+\tilde{s}\,e^{\,\xi})^{\chi_-+b}}\,,
 \eb
 where we have introduced the notation
 \be\label{sst}
 s=-\frac{r\,e^{\,\mathrm{Re}\,\tilde{\theta}_2}+r'}{1+rr'\,e^{\,\mathrm{Re}\,\tilde{\theta}_2}}
 \;e^{i\varphi'},\qquad
 \tilde{s}=-\frac{r\,e^{-\mathrm{Re}\,\tilde{\theta}_2}+r'}{1+rr'\,e^{-\mathrm{Re}\,\tilde{\theta}_2}}
 \;e^{-i\varphi'}.
 \eb
 Contour $C_{\xi}$ denotes the horizontal line segment in the
 $\xi$-plane, joining the points $\ln|s|+i\varphi'$ and
 $-\ln|\tilde{s}|+i\varphi'$. Rewriting (\ref{interm3}) in terms
 of ordinary integrals and using (\ref{sst}), one gets
 a relation analogous to (\ref{interm2}):
 \begin{eqnarray}
 \nonumber F(z,z',\tilde{\theta}_2)&=&\frac{1}{4\pi}\left(\frac{1+rr'\,e^{\,\mathrm{Re}\,\tilde{\theta}_2}
 }{1+rr'\,e^{-\mathrm{Re}\,\tilde{\theta}_2}}\right)^b\left(1-s\tilde{s}\right)^{\chi_-}
 \times \\
 \nonumber &\;&\times\int\limits_{\frac12\ln|s\tilde{s}|}^{-\frac12\ln|s\tilde{s}|}
 \frac{d\xi}{\left(1-{|s\tilde{s}|}^{1/2}\,e^{-\xi}\right)^{\chi_--b}
 \left(1-{|s\tilde{s}|}^{1/2}\,e^{\,\xi}\right)^{\chi_-+b}}=\\
 \label{need2}&=&\left(\frac{1+rr'\,e^{\,\mathrm{Re}\,\tilde{\theta}_2}
 }{1+rr'\,e^{-\mathrm{Re}\,\tilde{\theta}_2}}\right)^b\zeta\Bigl(|s\tilde{s}|\Bigr).
 \end{eqnarray}
 Finally, combining the last formula with (\ref{need1}) and
 (\ref{sst}), we find the relations (\ref{deltaans})--(\ref{vrr}).
 The same result can be obtained from the representation (\ref{delta2})
 in a completely analogous manner.

\renewcommand{\theequation}{B.\arabic{equation}}
\setcounter{equation}{0}

 \section*{Appendix B}
 In order to integrate the function $\Delta_k(t)$ over the
 hyperbolic disk, consider the following identity:
 \begin{eqnarray*}
 &\;& \left(\frac{1+te^{-\theta}}{1+te^{\;\theta}\;}\right)^b\zeta
 \left(\frac{2t(1+\cosh\theta)}{1+t^2+2t \cosh \theta}\right)=
 \\
 &=&\frac{(1-t)^{2\chi}\;}{4\pi}\int\limits_{-\infty}^{\infty}\frac{d\theta'}{
 \left[(1+e^{-\theta})(1+e^{-\theta'})-(1-t)e^{-\theta}\right]^{\chi-b}
 }\times\\
 &\times&\frac{1}{\left[(1+e^{\theta})(1+e^{\theta'})-
 (1-t)(e^{\theta}+e^{\theta'}+e^{\theta+\theta'})\right]^{\chi+b}}
 \,.
 \end{eqnarray*}
 Substituting this relation into the integral $\displaystyle \int\nolimits_D
 d\mu\;\Delta_k(t)$, integrating over the angle $\varphi$ and
 introducing instead of $t$ a new variable $s=1-t$, one obtains
 \be\label{integrald}
 \int\nolimits_D
 d\mu\;\Delta_k(t)=R^{\,2} \;
 \frac{\sin\pi\nu}{4\pi}\int\nolimits_{-\infty}^{\infty}d\theta
 \int\nolimits_{-\infty}^{\infty}d\theta'\;\frac{\;e^{\,(1+\nu)\theta}}{1+e^{\,\theta}}\,
 \mathcal{F}(\theta,\theta'),
 \eb
 \ben
 \mathcal{F}(\theta,\theta')=\int\nolimits_{0}^1 \frac{s^{2\chi-2}\;ds}{
 \left[(1+e^{-\theta})(1+e^{-\theta'})-se^{-\theta}\right]^{\chi-b}
 \left[(1+e^{\theta})(1+e^{\theta'})-
 s(e^{\theta}+e^{\theta'}+e^{\theta+\theta'})\right]^{\chi+b}}\,.
 \ebn
 After further change of variable $\displaystyle s\rightarrow u=
 \frac{s}{s+(1+e^{\theta})(1+e^{\theta'})(1-s)}$ the function $ \mathcal{F}(\theta,\theta')$
 can be rewritten as
 \ben
 \mathcal{F}(\theta,\theta')=\frac{1}{(1+e^{\theta})(1+e^{\theta'})}
 \int\nolimits_0^1\frac{u^{2\chi-2}du}{\left[e^{-\theta-\theta'}
 +(1+e^{-\theta'})u\right]^{\chi-b}}\,.
 \ebn
 Integrating once by parts, one finds
 \begin{eqnarray}
 \label{ftt1}\mathcal{F}(\theta,\theta')&=&
 \frac{\quad \left[ 1+e^{-\theta'}+e^{-\theta-\theta'}\right]^{-(\chi-b)}
 }{(2\chi-1)(1+e^{\theta})(1+e^{\theta'})}+\\
 \label{ftt2}&\;&+\frac{(\chi-b)e^{-\theta'}}{(2\chi-1)(1+e^{\theta})}
 \int\nolimits_0^1 \frac{u^{2\chi-1}du}{\left[e^{-\theta-\theta'}
 +(1+e^{-\theta'})u\right]^{\chi-b+1}}\,.
 \end{eqnarray}

 Let us consider the contribution of the term (\ref{ftt1}) to the integral
 (\ref{integrald}). Introducing instead
 of $\theta'$ a new variable $\displaystyle
 v=[1+e^{-\theta'}+e^{-\theta-\theta'}]^{-1}$,  we
 may write
 \be\label{auxres1}
 \int\nolimits_{-\infty}^{\infty}d\theta
 \int\nolimits_{-\infty}^{\infty}d\theta'\;\frac{\;e^{\,(1+\nu)\theta}}{1+e^{\,\theta}}\,
 \mathcal{F}_1(\theta,\theta')=\frac{1}{2\chi-1}\int\nolimits_0^1
 dv\int\nolimits_{-\infty}^{\infty}d\theta\;\frac{
 e^{\,(1+\nu)\theta}}{(1+e^{\,\theta})^2}\;\frac{v^{\chi-b-1}}{1+ve^{-\theta}}\,.
 \eb
 In the contribution of the term (\ref{ftt2}), we first perform the integration over
 $\theta'$, then replace $\theta\rightarrow-\theta$ and finally
 exchange the order of integration over $\theta$ and $u$. The
 result looks quite similar to (\ref{auxres1}):
 \be\label{auxres2}
 \int\nolimits_{-\infty}^{\infty}d\theta
 \int\nolimits_{-\infty}^{\infty}d\theta'\;\frac{\;e^{\,(1+\nu)\theta}}{1+e^{\,\theta}}\,
 \mathcal{F}_2(\theta,\theta')=\frac{1}{2\chi-1}\int\nolimits_0^1
 du\int\nolimits_{-\infty}^{\infty}d\theta\;\frac{
 e^{-\nu\theta}}{(1+e^{\,\theta})^2}\;\frac{u^{\chi+b-1}}{1+ue^{-\theta}}\,.
 \eb
 The integrals over $\theta$ in (\ref{auxres1})--(\ref{auxres2})
 can be easily calculated by residues. For example, one obtains
 \ben
 \frac{\sin\pi\nu}{\pi}\int\nolimits_{-\infty}^{\infty}d\theta\;\frac{
 e^{\,(1+\nu)\theta}}{(1+e^{\,\theta})^2(1+ve^{-\theta})}=
 \frac{v^{1+\nu}-1+(1+\nu)(1-v)}{(1-v)^2}\,.
 \ebn
 Subsequent integration over $v$ and $u$ leads to the final
 result:
 \begin{eqnarray*}
  \int\nolimits_D
 d\mu\;\Delta_k(t)&=&-\frac{R^{\,2}}{4(2\chi-1)} \;
 \left\{(\chi-b+\nu)\Bigl[\psi(\chi-b)-\psi(\chi-b+\nu+1)\Bigr]+\right.\\
 \nonumber&\;& \left. +\;
 (\chi+b-\nu-1)\Bigl[\psi(\chi+b)-\psi(\chi+b-1-\nu)\Bigr]\right\},
 \end{eqnarray*}
 where $\psi(x)$ denotes the digamma function.


\begin{thebibliography}{100}
 \bibitem{abst}
 M. Abramowitz, I. A. Stegun, \textit{Handbook of mathematical
 functions}, New York, Dover, (1965).
 \bibitem{adami}
 R. Adami, A. Teta, \textit{On the Aharonov-Bohm hamiltonian},
 Letts. Math. Phys.~\textbf{43}, (1998), 43--54; preprint
 \texttt{quant-ph/9702048}.
 \bibitem{aeg}
 S. A. Albeverio, P. Exner, V. A. Geyler, \textit{
 Geometric phase related to point-interaction transport on a
 magnetic Lobachevsky plane}, Letts. Math. Phys. \textbf{55}, (2001), 9--16;
 preprint \texttt{math-ph/0008031}.
  \bibitem{albeverio} S. Albeverio, F. Gesztesy, R. H{\o}egh-Krohn,
 H. Holden, \textit{Solvable models in quantum mechanics},
 Springer, (1988).
 \bibitem{bruning}
 J.~Br\"uning, V. A. Geyler, \textit{Gauge-periodic point
 perturbations on the Lobachevsky plane}, Theor.~\&~Math.
 Phys.~\textbf{119}, (1999), 687--697.
 \bibitem{bgm}
 D. V. Bulaev, V. A. Geyler, V. A. Margulis, \textit{Quantum Hall
 effect on the Lobachevsky plane}, Physica~\textbf{B337}, (2003),
 180--185.
 \bibitem{bulaev}
 D. V. Bulaev, V. A. Geyler, V. A. Margulis, \textit{Effect of surface curvature on magnetic
 moment and persistent currents in two-dimensional quantum rings and dots}, Phys. Rev.~\textbf{B69},
 (2004), 195313; preprint \texttt{cond-mat/0308500}.
 \bibitem{comtet_houston}
 A. Comtet, P. J. Houston, \textit{Effective action on the
 hyperbolic plane in a constant external field}, J. Math.
 Phys.~\textbf{26}, (1985), 185--191.
 \bibitem{comtet}
 A. Comtet, \textit{On the Landau levels on the
 hyperbolic plane}, Ann. Phys. \textbf{173}, (1987), 185--209.
 \bibitem{comtet_jpa}
 A. Comtet, Y. Georgelin, S. Ouvry, \textit{Statistical aspects of the anyon
 model}, J. Phys.~\textbf{A22}, (1989), 3917--3925.
  \bibitem{stovicek1}
 L. Dabrowski, P. Stovicek, \textit{Aharonov-Bohm effect with
 $\delta$-type interaction}, J. Math. Phys.~\textbf{39}, (1998),
 47--62; preprint \texttt{physics/9612014}.
 \bibitem{DeMicheli}
 E. De Micheli, I. Scorza, G. A. Viano, \textit{Hyperbolic
 geometrical optics: hyperbolic glass}, J. Math.
 Phys.~\textbf{47}, (2006), 023503.
 \bibitem{desbois1}
 J. Desbois, C. Furtlehner, S. Ouvry, \textit{Random magnetic impurities and the $\delta$-impurity
 problem}, J. Phys.~I France~\textbf{6}, (1996), 641-648; preprint
 \texttt{cond-mat/9412076}.
  \bibitem{desbois2}
 J. Desbois, C. Furtlehner, S. Ouvry, \textit{Random magnetic impurities and the
 Landau problem}, Nucl. Phys.~\textbf{B453}, (1995), 759-776; preprint
 \texttt{cond-mat/9509105}.
 \bibitem{desbois3}
 J. Desbois, S. Ouvry, C. Texier,
 \textit{Persistent currents and magnetization in two-dimensional magnetic
 quantum systems}, Nucl. Phys.~\textbf{B528}, (1998), 727-745;
 preprint \texttt{cond-mat/9801106}.
  \bibitem{desousa}
 Ph. de Sousa Gerbert, \textit{Fermions in an
 Aharonov-Bohm field and cosmic strings}, Phys. Rev.~\textbf{D40},
 (1989), 1346--1349.
 \bibitem{doyon}
 B. Doyon, \textit{Two-point correlation
 functions of scaling fields in the Dirac theory on the Poincare
 disk}, Nucl. Phys. \textbf{B675}, (2003), 607--630; preprint
 \texttt{hep-th/0304190}.
 \bibitem{doyon2}
 B. Doyon, \textit{Form factors of Ising spin and
 disorder fields on the Poincar\'e disk}, J. Phys. \textbf{A37},
 (2004), 359--370.
 \bibitem{doyon_fonseca} B. Doyon, P. Fonseca, \textit{Ising field
 theory on a pseudosphere}, J. Stat. Mech.~\textbf{0407}, (2004),
 P002; preprint \texttt{hep-th/0404136}.
 \bibitem{elstrodt}
 J. Elstrodt, \textit{Die resolvente zum Eigenwertproblem der
 automorphen Formen in der hyperbolischen Ebene}, Math. Ann.
 \textbf{203}, (1973), 295--330.
 \bibitem{stovicek2}
 P. Exner, P. Stovicek, P. Vytras, \textit{Generalized boundary conditions
 for the Aha\-ro\-nov-Bohm effect combined with a homogeneous magnetic
 field}, J.\- Math.\-Phys. \textbf{43}, (2002), 2151--2168; preprint
 \texttt{quant-ph/0111050}.
  \bibitem{falomir}
 H. Falomir, P. A. G. Pisani, \textit{Hamiltonian self-adjoint
 extensions for (2+1)-di\-men\-sional Dirac particles}, J. Phys. \textbf{34}, (2001),
 4143-4154;
 preprint \texttt{math-ph/0009008}.
 \bibitem{grosche_h}
 C. Grosche, \textit{Path integration on the hyperbolic plane with
 a magnetic field}, Ann. Phys.~\textbf{201}, (1990), 258--284.
 \bibitem{grosche}
 C. Grosche, \textit{On the path integral treatment for an Aharonov-Bohm field
 on the hyperbolic plane}, Int. J. Theor. Phys.~\textbf{38}, (1999),
 955--969; preprint \texttt{quant-ph/9808060}.
 \bibitem{grosche_c}
 C. Grosche, \textit{The path integral for the Kepler problem on
 the pseudosphere}, Ann. Phys.~\textbf{204}, (1990), 208--222.
 \bibitem{kuperin} Yu. A. Kuperin, R. V. Romanov, H. E. Rudin,
 \textit{Scattering on the hyperbolic plane in the Aharonov-Bohm gauge field},
 Letts. Math. Phys.~\textbf{31}, (1994), 271--278.
 \bibitem{marino}
 E. C. Marino, B. Schroer, J. A. Swieca, \textit{Euclidean
 functional integral approach for disorder variables and kinks},
 Nucl. Phys.~\textbf{B200}, (1982), 473--497.
 \bibitem{narayanan}
 R. Narayanan, C. A. Tracy, \textit{Holonomic quantum field theory
 of bosons in the Poincar\'e disk and the zero curvature limit},
 Nucl. Phys. \textbf{B340}, (1990), 568--594.
 \bibitem{nouicer}
 Kh. Nouicer, L. Chetouani, \textit{Path integral for relativistic Aharonov-Bohm-Coulomb
 system on the pseudosphere}, J. Math. Phys.~\textbf{42}, (2001),
 1053--1065.
 \bibitem{pacific}
 J. Palmer, \textit{Tau functions for the Dirac
 operator in the Euclidean plane}, Paci\-fic J. Math.~\textbf{160},
 (1993), 259--342.
 \bibitem{beatty} J. Palmer, M. Beatty, C. A. Tracy, \textit{Tau functions
 for the Dirac operator on the Poincar\'e disk},
 Comm. Math. Phys.~\textbf{165}, (1994), 97--173; preprint
 \texttt{hep-th/9309017}.
 \bibitem{pt}
 J. Palmer, C. A. Tracy, \textit{Monodromy preserving deformation
 of the Dirac operator acting on the hyperbolic plane}, in ``Mathematics
 of Nonlinear Science: proceedings of an AMS special session held January~11--14, 1989'',
 ed. M. S. Berger, Contemporary Mathematics~\textbf{108}, (1990),
 119--131.
 \bibitem{sitenko} Yu. A. Sitenko, V. M. Gorkavenko,
 \textit{Induced quantum numbers of a magnetic vortex at non-zero
 temperature}, Nucl. Phys.~\textbf{B714}, (2005), 217--255; preprint \texttt{hep-th/0410091}.
\end{thebibliography}
\end{document}